\documentclass[cits]{PoS}

\title{Discovery of VHE $\gamma$-ray emission from the W49 region with H.E.S.S.}

\ShortTitle{VHE $\gamma$-ray emission from the W49 region with H.E.S.S.}

\author{Francois Brun$^1$, Mathieu de Naurois$^1$, Werner Hofmann$^2$, Svenja Carrigan$^2$, Arache Djannati-Ata\"i$^3$ and Stefan Ohm$^{245}$ for the H.E.S.S. Collaboration$^6$\\%
  Leprince-Ringuet, Ecole Polytechnique, CNRS/IN2P3, Palaiseau,
  France\\ \llap{$^2$} Max Planck Institut f\"ur Kernphysik,
  Heidelberg, Germany\\ \llap{$^3$} APC - AstroParticule Cosmologie,
  CNRS/IN2P3 - Univ Paris 7 - Observatoire de Paris - CEA, Paris,
  France\\ \llap{$^4$} School of Physics \& Astronomy, University of
  Leeds, UK\\ \llap{$^5$} Department of Physics \& Astronomy,
  University of Leicester, UK\\ \llap{$^6$}
  http://www.mpi-hd.mpg.de/hfm/HESS\\ E-mail:
  \email{francois.brun@llr.in2p3.fr}}

\abstract{The W49 region hosts two bright radio sources: the star
  forming region W49A and the supernova remnant W49B. The
  $10^{6}~M_\odot$ Giant Molecular Cloud W49A is one of the most
  luminous giant radio H{\sc ii} regions in our Galaxy and hosts
  several active, high-mass star formation sites. The mixed-morphology
  supernova remnant W49B has one of the highest surface brightness in
  radio of all the SNRs of this class in our Galaxy and is one of the
  brightest ejecta-dominated SNRs in X-rays. Infrared observations
  evidenced that W49B is interacting with molecular clouds and Fermi
  recently reported the detection of a coincident bright, high-energy
  gamma-ray source. Observations by the H.E.S.S. telescope array
  resulted in the significant detection of VHE gamma-ray emission from
  the W49 region, compatible with VHE emission from the SNR W49B. The
  results, the morphology and the origin of the VHE emission are
  presented in the multi-wavelength context and the implications on
  the origin of the signal are discussed.}

\FullConference{25th Texas Symposium on Relativistic Astrophysics \\
		 December 6-10, 2010\\
		 Heidelberg, Germany}

\begin{document}

\section{The W49 Region}

The W49 region is a prime candidate to observe with ground-based
Cherenkov telescopes such as H.E.S.S. since it hosts a star forming
region (W49A) and a mixed morphology supernova remnant interacting
with molecular clouds (W49B).

W49A is one of the most luminous giant H{\sc ii} region in the Galaxy
\cite{Smith}. In the core ($\sim 15$ pc) of this $10^6~M_\odot$ Giant
Molecular Cloud of 100 pc in total extension \cite{Simon}, $\sim 30$
ultra-compact H{\sc ii} regions, each hosting at least one massive
star (earlier than B3) are resolved in radio \cite{dePree}. From the
proper motion of the strong $H_2O$ masers it hosts, the distance of
W49A is estimated to be $11.4\pm1.2~\mathrm{kpc}$ \cite{Gwinn}.

The progenitor of W49B is thought to be a super-massive star that
created a wind-blown bubble in a dense molecular cloud in which the
explosion occured \cite{Keohane} as revealed by IR and X-ray
observations. The detection of Mid-IR lines from shocked molecular
hydrogen is an evidence that W49B is interacting with molecular clouds
\cite{Reach}. From H{\sc i} absorption analyses, its distance was
estimated to be $\sim 8~\mathrm{kpc}$ \cite{Radha}. More recent VLA
observations show that W49B could be associated with W49A
\cite{Brogan}, extending the range of possible distances for this
object ($8~\mathrm{kpc} < D < 12~\mathrm{kpc}$). Its age is estimated
to be between $1~\mathrm{kyr}$ and $4~\mathrm{kyrs}$ \cite{Pye}
\cite{Hwang}. In Radio, the supernova remnant shell is resolved with a
size of $\sim 4'$. W49B is also detected by the Fermi-LAT at a level
of $38 \sigma$ with 17 months of data \cite{AbdoW49}.

The discovery of VHE $\gamma$-rays from the W49 region is reported in
the next section. These preliminary results are then discussed.

\section{H.E.S.S. Observations and Analysis Results}

H.E.S.S. is an array of four 13 m diameter imaging Cherenkov
telescopes situated in the Khomas Highlands in Namibia at an altitude
of 1800 m above sea level (see e.g. \cite{Bern}, \cite{Funk}).  The
standard H.E.S.S. run selection procedure was used to select
observations taken under good weather conditions. This resulted in a
dataset comprising 60 hours of observations (live time) on W49B and
W49A. Data were analysed using the \emph{Model Analysis} as described
in \cite{Naurois}. This analysis was performed on W49A and W49B, using
the standard cuts which include a minimum charge of $60$
photoelectrons resulting in an energy threshold of $\sim
260~\mathrm{GeV}$. The analysis regions were defined a-priori as
circles of $0.1^\circ$ centered on the nominal position of W49A ($l =
43.17^\circ, b = 0.0^\circ$) and W49B ($l = 43.27^\circ, b =
-0.19^\circ$). The results presented below were also confirmed by
independent analyses such as those described in \cite{Ohm_MVA} or
\cite{Becherini_MVA}.

Figure 1 shows the resulting excess map smoothed to the H.E.S.S. Point
Spread Function (PSF) ($68\%$ containment radius $R_{68} =
0.066^\circ$).  An excess of $191$ VHE $\gamma$-rays is detected
towards W49B by H.E.S.S. with a statistical significance of $8.8
\sigma$ using an integration radius of $0.1^{\circ}$.  An excess of
VHE $\gamma$-rays is also detected in the direction of W49A with a
significance of more than $4.4 \sigma$.  The best fit position of the
TeV emission is found to be ($l = 43.258^\circ\pm0.008^\circ, b =
-0.188^\circ\pm0.01^\circ$) assuming point-like emission. As shown on
Figure 2, this is well coincident with the brightest radio part of the
W49B remnant and with the GeV emission fitted position ($l =
43.251^\circ - b = -0.168^\circ$, with an error radius of
$0.024^\circ$ at $95\%$ C.L.\cite{AbdoW49}). 

The TeV excess visible towards W49A is in good coincidence with the
densest part of the molecular cloud as observed by the $^{13}CO$
Galactic Ring Survey \cite{Jackson} as can be seen on Figure 2.

\begin{figure}[tbh!]
  \begin{center}
  \includegraphics[width=.6\textwidth]{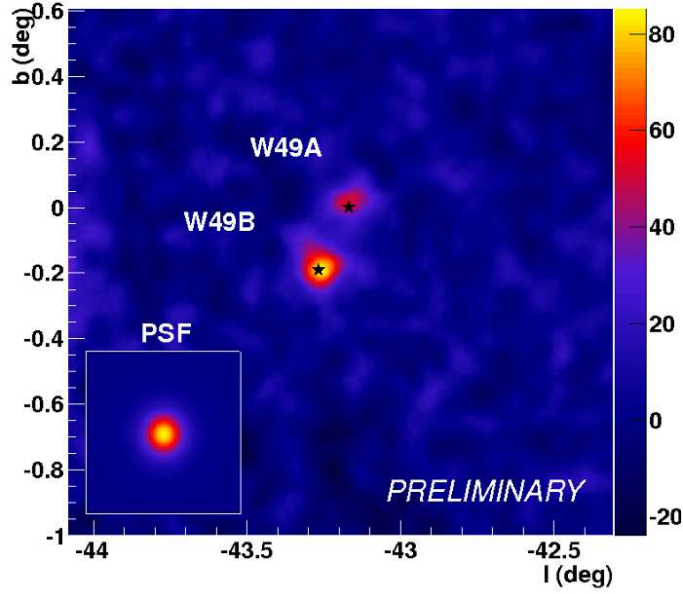}
  \caption{H.E.S.S. excess map of the W49 region obtained with the
    Model Analysis. The map is smoothed to the H.E.S.S. PSF shown in
    the caption. The stars marks W49B and W49A nominal positions.}
  \end{center}
\end{figure}

\begin{figure}[tbh!]
  \begin{center}
    \includegraphics[width=.9\textwidth]{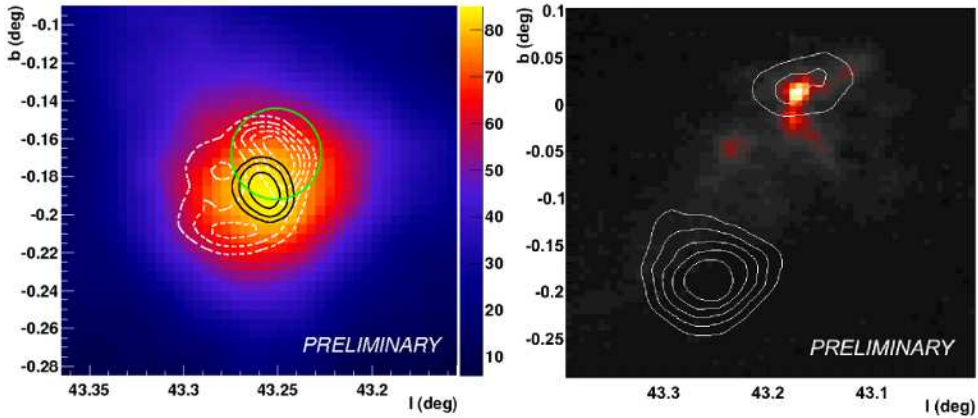}
    \caption{Left: Detail of Figure 1 centered on W49B. The black
      contours are the error contours at 68\%, 95\% and 99\% of the
      fitted position assuming point-like emission. The green circle
      is the Fermi-LAT fitted position at 95\% C.L. The white contours
      show the radio emission as seen by NVSS. Right: Integrated Map
      of the $^{13}CO(J=1-0)$ Galactic Ring Survey between $v_{LSR} =
      0~\mathrm{km/s}$ and $20~\mathrm{km/s}$. This velocity range
      corresponds to the distance to W49A. The white contours are from
      the H.E.S.S. excess map shown in Fig. 1.}
  \end{center}
\end{figure}

The differential energy spectrum of the VHE $\gamma$-ray emission
towards W49B was derived above the energy threshold of $260
\mathrm{GeV}$ selecting the events inside a circular region of
$0.1^{\circ}$ around the supernova remnant nominal position.  The
spectrum obtained for W49B is well described ($\chi^2/dof = 39.6/38$)
by a power-law model defined as $dN/dE = N_0(E/1TeV)^{-\Gamma}$ with
$\Gamma = 3.1\pm0.3_{stat}\pm0.2_{syst}$ and $N_0 =
2.3\pm0.4_{stat}\pm0.6_{syst}
10^{-13}~\mathrm{cm^{-2}}\mathrm{s^{-1}}\mathrm{TeV^{-1}}$. This
corresponds to an integral flux above $1 \mathrm{TeV}$ of
$1.1\pm0.3_{stat}\pm0.3_{syst} 10^{-13}~\mathrm{cm^{-2}}
\mathrm{s^{-1}}$, equivalent to $\sim 0.5\%$ of the Crab nebula flux
above the same energy.  As can be seen on Figure 3, the GeV
\cite{AbdoW49} and TeV gamma-ray spectra are in remarkably good
agreement.

\begin{figure}
  \begin{center}
  \includegraphics[width=.6\textwidth]{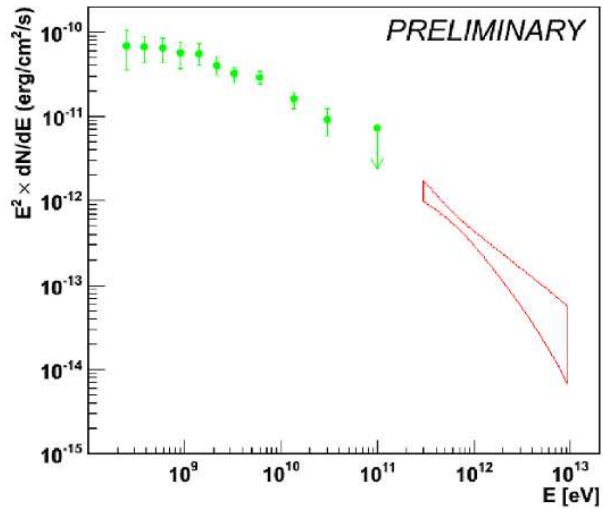}
  \caption{Combined GeV and TeV energy spectrum obtained by the
    Fermi-LAT (green) and H.E.S.S. (red). The H.E.S.S. spectrum was
    extracted from a circular region of $0.1^\circ$ around the nominal
    position of W49B.}
  \end{center}
\end{figure}

\section{Interpretations}

\subsubsection*{W49B}

The most straight forward interpretations for the origin of the signal
are particle acceleration either by a pulsar or by the SNR shock. No
observations suggest the presence of a pulsar or pulsar wind nebula in
W49B.  The spatial coincidence of the TeV emission with the brightest
part of the radio shell and GeV emission points toward emission from
particle accelerated at the shock as predicted, for instance by the
diffusive shock acceleration theory (DSA) \cite[e.g.]{Drury83}

Since the shock is observed to be interacting with the molecular cloud
in which the supernova exploded, very-high energy $\gamma$-ray
emissions from the decay of $\pi^0$ mesons is expected to be enhanced
proportionnaly to the target mass. Furthermore, the large GeV
gamma-ray luminosity \cite{AbdoW49} of $\sim
10^{36}~\mathrm{erg}~\mathrm{s^{-1}}$ seems to be difficult to explain
with IC scattering only.  The detection of W49B at GeV and TeV
emission is therefore a rather compelling argument in favour of a
hadronic nature of the accelerated particles.  More detailed studies
are in progress to understand and constrain the emission processes in
W49B.

\subsubsection*{W49A}

Star forming regions are potential acceleration sites of VHE
particles. This can, for instance, occur at the shocks created by the
strong winds of the numerous massive stars they generally host.
Recently, the presence of two expanding shells as well as gas
ejections were found in W49A \cite{Peng}. The shells seem to have a
common origin in the cloud core and a total kinetic energy of $\sim
10^{49}~\mathrm{ergs}$. The gas ejections are likely to have the same
origin as the expanding shells and a total energy of $\sim
10^{50}~\mathrm{ergs}$.

\section{Conclusion}

The W49 region was observed by the H.E.S.S. telescope array, yielding
$\sim 60 \mathrm{h}$ of good quality data. This led to the significant
detection of TeV gamma-ray emission coincident with the supernova
remnant W49B at a significance level of $8.8 \sigma$. The position of
the emission is compatible with the brightest part of the radio
emission from the SNR as well as with the GeV emission. Energy spectra
in the GeV and TeV bands are in very good agreement. Given the very
high GeV luminosity and the fact that the SNR is interacting with
dense material, a hadronic scenario is favored.

These observations also resulted in evidence for gamma-ray emission in
the direction of the star forming region W49A.  Analyses are still
ongoing in order to confirm this promising preliminary result.

\acknowledgments

The support of the Namibian authorities and of the University of
Namibia in facilitating the construction and operation of H.E.S.S. is
gratefully acknowledged, as is the support by the German Ministry for
Education and Research (BMBF), the Max-Planck-Society, the French
Ministry for Research, the CNRS-IN2P3 and the Astroparticle
Interdisciplinary Programme of the CNRS, the U.K. Particle Physics and
Astronomy Research Council (PPARC), the IPNP of the Charles
University, the South African Department of Science and Technology and
National Research Foundation, and by the University of Namibia. We
appreciate the excellent work of the technical support staff in
Berlin, Durham, Hamburg, Heidelberg, Palaiseau, Paris, Saclay, and in
Namibia in the construction and operation of the equipment.

This publication makes use of molecular line data from the Boston
University-FCRAO Galactic Ring Survey (GRS). The GRS is a joint
project of Boston University and Five College Radio Astronomy
Observatory, funded by the National Science Foundation under grants
AST-9800334, AST-0098562, AST-0100793, AST-0228993, \& AST-0507657.

\end{document}